# Fundamental limit of light trapping in grating structures

**Zongfu Yu\*, Aaswath Raman and Shanhui Fan[+]**

*Ginzton Lab, Stanford University, Stanford, CA, 94305 USA*
*\*zfyu@stanford.edu, [+]shanhui@stanford.edu*

**Abstract:** We use a rigorous electromagnetic approach to analyze the fundamental limit of light-trapping enhancement in grating structures. This limit can exceed the bulk limit of $4n^2$, but has significant angular dependency. We explicitly show that 2D gratings provide more enhancement than 1D gratings. We also show the effects of the grating profile's symmetry on the absorption enhancement limit. Numerical simulations are applied to support the theory. Our findings provide general guidance for the design of grating structures for light-trapping solar cells.

## 1. Introduction

Light-trapping schemes can be used to enhance absorption in photovoltaic (PV) cells. These schemes help to increase cell efficiency and also reduce production cost. Understanding the absorption enhancement that is maximally achievable by light trapping schemes is thus of fundamental importance in solar cell research. In a homogeneous bulk cell with a back reflection mirror, light trapping can be accomplished by introducing random roughness on the surface of the cell. For these cells, it is known that, with respect to the single-pass absorption of the film, the maximum enhancement factor achievable by light trapping schemes is $4n^2/\sin^2(\theta)$, where $n$ is the refractive index of the material and $\theta$ is half of apex angle of the absorption cone [1-4]. For a Lambertian cell, with an isotropic response and hence $\theta = 90^o$, the maximum absorption enhancement factor reduces to $4n^2$. This limit of $4n^2/\sin^2(\theta)$ will be referred to as the "bulk limit" in this paper.

The bulk limit is derived with the following assumptions [1-4]:

(a) One assumes a film with a thickness that is much larger than the wavelength. Conventional silicon cells, for example, have a thickness of hundreds of microns.

(b) The light trapping involves the use of either a random roughness, or a grating structure in which the periodicity is much larger than the wavelength of light. It has been shown in [1-3] that a grating with periodicity much larger than the wavelength in fact provides a maximum enhancement factor that is the same as a structure with random roughness.

(c) The single-pass absorption of the film is negligible. Historically, the concept of light trapping was first developed to enhance absorption in the band edge region of a bulk crystalline silicon cell, where the absorption coefficient is typically on the order of $10^{-2}$ to $10^{-1}$ cm$^{-1}$. The bulk limit can only be achieved when the absorber is sufficiently weak such that, even after multiplying by the maximum enhancement factor, the absorption of the film remains relatively small. The relevant figure of merit in such a weak absorber is thus the enhancement ratio rather than the film's absorption coefficient. The maximum absorption enhancement factor decreases when the single-pass absorption of the film becomes significant [2, 5]. Nevertheless, many insights on light trapping, as obtained from the study of weak absorbers, are generally applicable. Thus a significant number of theoretical works have focused on weak absorbers [2, 6-8].

The combination of the first two assumptions above (a relatively thick film and large periodicity) ensures that wave effects, including interference and diffraction, become weak, at least in a spectrally averaged sense. As a result, the bulk limit can be derived by a ray-tracing argument[2, 5]. Therefore, one may consider the bulk limit as a "ray-optics" limit, even though it needs to be emphasized that the bulk limit is correct from a rigorous electromagnetic wave perspective [2, 9], .

The bulk limit is not strictly applicable when the wave effects of light become prominent. This occurs when the film thickness is comparable or smaller than the wavelength of interest, or when one uses a periodic grating with periodicity comparable to the wavelength of light. Understanding light trapping in this new "wave-optics" regime is of great practical importance for many on-going experimental works on nanostructure-enhanced solar cells [10-22], and has also been the subject of many theoretical and numerical studies [7-8, 23-32]. Most of these theoretical and numerical studies, while providing specific results on particular structures, do not discuss the general upper limit of light absorption enhancement in the "wave-optics" regime. There have been a few attempts [8, 33] aiming to calculate the upper limit of the enhancement ratio in this wave regime by incoherently summing the scattering waves. Such an incoherent summation calculation is in its spirit similar to the ray optics method where phase information is neglected. It is however difficult to establish the general validity of these calculations in the "wave-optics" regime.

In Ref. [9], we developed a statistical temporal coupled mode theory formalism, which allows one to calculate the maximum light absorption enhancement ratio in the "wave-optics" regime from a rigorous electromagnetic perspective. Using this theory, we considered a structure where the film thickness is at a deep-subwavelength scale, which violates assumption (a) above, and showed that the maximum absorption enhancement ratio can significantly exceed the conventional bulk limit.

In this paper, we apply the statistical temporal coupled mode theory of Ref. [9] to study the case of light trapping in grating structures with wavelength-scale periodicity. The use of wavelength scale periodicity violates assumption (b) in the conventional theory, necessitating our new approach. To specifically probe the effect of grating periodicity, we will maintain all other assumptions made in the conventional theory, i.e. we will be considering a weakly absorbing film with a thickness that is at least a few wavelengths thick. We show that in a grating structure with wavelength scale periodicity, the limit of absorption enhancement is significantly different from the bulk limit. We further show that the theoretical analysis compares well with direct numerical simulations.

The paper is organized as follows: In section 2, we describe the grating structures. In section 3, we review the theory used to derive the absorption enhancement limit. In sections 4 and 5, we describe theories for one-dimensional (1D) and two-dimensional (2D) gratings, respectively. In section 6, we discuss the regime of applicability for the theory. Finally, we conclude in Section 7.

## 2. The model grating structures

As a concrete example to illustrate the effect of gratings in absorption enhancement, in this paper we consider the class of structures shown in Fig. 1a, where a shallow grating is introduced above a uniform active layer. The whole structure is placed on a reflecting mirror (Fig. 1a). The thickness of the active layer is chosen to be a few wavelengths thick. (Here we choose a film with a thickness of 3 μm). Understanding absorption enhancement in films with such thickness is of importance in practice. For example, crystalline silicon cells are typically tens and even hundreds of micro-meter thick due to the materials' weak absorption close to the band-edge.

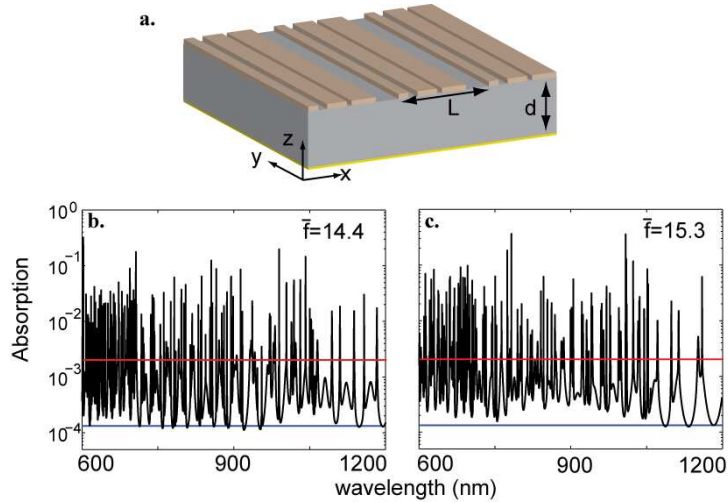

Fig. 1. a) Grating structure with periodicity $L$=600nm. Brown ribbons are non-absorptive dielectric medium with index $n^2$ =12.5. The widths of three air slots are 4%, 12%, and 24% of the period respectively. The absorptive film (grey) is $d$=3000nm thick. The whole structure is placed on a perfect mirror (yellow). b) and c) Absorption spectra obtained by numerical simulations for s (b) and p (c) polarized incident light. Light is incident from the normal direction. Red lines represent spectrally-averaged absorption in the presence of the grating and blue lines represent single-pass absorption in the absence of the grating.

Such a film supports guided optical modes. In the limit where the absorption of the active layer is weak, these guided modes typically have a propagation distance along the film that is much longer than the thickness of the film. Light trapping is accomplished by coupling incident plane-waves into these guided modes with a grating (Fig. 1a). As long as the periodicity $L$ of the grating is chosen to be sufficiently large, i.e. at least comparable to the free-space wavelength of the incident light, each incident plane-wave can couple into at least one guided mode. By the same argument, such a guided mode can couple to external plane-waves, creating a guided resonance [34].

As an illustration, Fig. 1b shows the absorption spectrum of the structure in Fig. 1a. The active material has a high refractive index ($n^2$ = 12.5) and a long absorption length of 22.5mm over the entire solar spectrum, which approximates silicon's properties at band-edge. In the absence of the grating, the single-pass absorption of the film is very weak. (Blue line, Fig. 1b). For the grating structure, we simulate its absorption properties by solving Maxwell's equations exactly using rigorous coupled wave analysis [34]. The resulting absorption spectrum consists of multiple peaks, each corresponding to a guided resonance. The absorption is strongly enhanced in the vicinity of each resonance. However, compared to the broad solar spectrum, each individual resonance has very narrow spectral width. Consequently, to enhance absorption over a substantial portion of the solar spectrum, one must rely upon a collection of these peaks.

Motivated by this observation, in the following sections we develop a statistical temporal coupled mode theory to describe the aggregate contributions from all resonances. Using this theory, we analytically calculate the theoretical limit of absorption enhancement in grating structures. We then compare these predictions with numerical simulations.

The aim of our paper is to establish an upper bound on absorption enhancement in such grating structures. The role of numerical simulations in such an effort therefore requires some elucidation. As a matter of principle, one cannot establish an upper bound using numerical simulations alone, since in order to do so, one would need to simulate an infinite number of structures. The numerical simulations, rather, are used here for two purposes. First, we will show that many of the structures simulated here in fact have absorption enhancement ratios that are comparable to the theoretical upper bound. Thus, the theoretical upper bound is in fact a very tight one. Secondly, the theory will show that different classes of grating structures have very different upper bounds on their enhancement ratios. For example, the theory will show that a 2D grating has an upper bound that is significantly higher than 1D gratings. Correspondingly, the numerical simulations will show that the actual enhancement factors in these different classes of structures are in fact also quite different. As a result, the insights that one obtains by the theoretical study are quite relevant in the actual physical structure.

## 3. Overview of the theory

In order to describe the upper limit of absorption enhancement in grating structures, here we briefly reproduce the relevant aspects of the statistical coupled-mode theory proposed in [9]. Since the grating structure enhances absorption through the aggregate contribution of a large number of resonances as seen in Fig. 1, we start by identifying the contribution of a single resonance to the total absorption over a broad spectrum. The behavior of an individual guided resonance, when excited by an incident plane wave, is described by the temporal coupled mode theory equation [35-36]:

$$\frac{d}{dt}a = (j\omega_0 - \frac{N\gamma_e + \gamma_i}{2})a + j\sqrt{\gamma_e}S \quad (1)$$

Here $a$ is the resonance amplitude, normalized such that $|a|^2$ is the energy per unit area in the film. $\omega_0$ is the resonance frequency. $\gamma_i$ is the intrinsic loss rate of the resonance due to material absorption. $S$ is the amplitude of the incident plane wave, with $|S|^2$ corresponding to its intensity. We refer to a plane wave that couples to the resonance as a *channel*. $\gamma_e$ is the leakage rate of the resonance to the channel that carries the incident wave. In general, the grating may phase-match the resonance to other plane-wave channels as well. We assume a total of $N$ such channels. Equivalent to the assumption of a Lambertian emission profile as made in [2], we further assume that the resonance leaks to each of the $N$ channels with the same rate $\gamma_e$. Under these assumptions, the absorption spectrum of the resonance is [9]

$$A(\omega) = \frac{\gamma_i \gamma_e}{(\omega - \omega_0)^2 + (\gamma_i + N\gamma_e)^2/4} \quad (2)$$

For light trapping purposes, the incident light spectrum is typically much wider than the linewidth of the resonance. When this is the case, we characterize the contribution of a single resonance to the total absorption by a *spectral cross-section*:

$$\sigma = \int_{-\infty}^{\infty} A(\omega)d\omega \quad (3)$$

Notice that spectral cross-section has a unit of frequency. For an incident spectrum with bandwidth $\Delta\omega \gg \sigma$, a resonance contributes $\sigma/\Delta\omega$ to the spectrally-averaged absorption coefficient. For a single resonance, from Eqs. (2) and (3), its spectral cross-section is:

$$\sigma = 2\pi \frac{\gamma_i}{N + \gamma_i/\gamma_e} \qquad (4)$$

which reaches a maximum value of

$$\sigma_{max} = \frac{2\pi\gamma_i}{N} \qquad (5)$$

in the *over-coupling* regime when $\gamma_e \gg \gamma_i$. We emphasize that the requirement to operate in the strongly over-coupling regime arises from the need to accomplish broad-band absorption enhancement.

We can now calculate the upper limit for absorption by a given medium, by summing over the maximal spectral cross-section of all resonances:

$$A = \frac{\sum \sigma_{max}}{\Delta\omega} = \frac{2\pi\gamma_i}{\Delta\omega} \frac{M}{N} \qquad (6)$$

where *M* is number of resonances within the frequency range $\Delta\omega$. In the over-coupling regime, the peak absorption from each resonance is in fact relatively small; therefore the total cross section can be obtained by summing over the contributions from individual resonances. In addition, we assume that the medium is weakly absorptive such that single-pass light absorption $\alpha d$ is negligible. Comparing the average absorption with the single-pass absorption, the upper limit of enhancement factor is

$$F = \frac{A}{\alpha d} = \frac{2\pi\gamma_i}{\alpha d \Delta\omega} \frac{M}{N} \qquad (7)$$

where $\alpha = n\gamma_i/c$ is the absorption coefficient.

Eq. (7) is the main result of the theory. We note that the maximal enhancement ratio is determined by both the properties of free space, in terms of the number of accessible plane wave channels *N*, as well as the film itself, in terms of the number of resonances *M*. In the case of films that are a few wavelengths thick, with only a shallow grating on top, as we consider here, the modification of the number of resonances over a broad range of frequencies is minimal. As a result, the main effect of the grating is to affect the number of accessible channels in free space. In the following sections, we will use Eq. (7) to calculate the light-trapping limit for 1D and 2D grating structures and to compare it to direct numerical simulations.

**4. 1D grating**

We define a 1D grating as structures that are uniform in one dimension, e.g. *y*-direction in Fig. 1a, and are periodic in the other dimension (*x*-direction in Fig. 1a) with a periodicity *L*. We consider incident light propagating in the *xz*-plane. Since the structure considered is two-dimensional, one can separately consider the response of the structures for the *s*-polarization, which has its electric field polarized along the *y*-direction, and for the *p*-polarization, which has its magnetic field polarized along the *y*-direction. In both the theoretical analysis and the numerical simulation shown in this Section, we will consider the two polarizations separately.

*4.1 Light trapping in grating structures without mirror symmetry*

We first consider the general case where the grating profile has no mirror symmetry along the *x*-direction (Fig. 1a). The effect of symmetry will be illustrated in Section 4.2.

For light incident from the normal direction, the periodicity results in the excitation of other plane waves with $k_x = 0, \pm 2\pi/L, \pm 4\pi/L, \cdots$ in the free-space above. Moreover, since these plane waves are propagating modes in air, one needs to have $k_x \leq k_0$, where $k_0$ is the wavevector of the incident light. These two requirements completely specify the number of channels available in *k*-space (Fig. 2a).

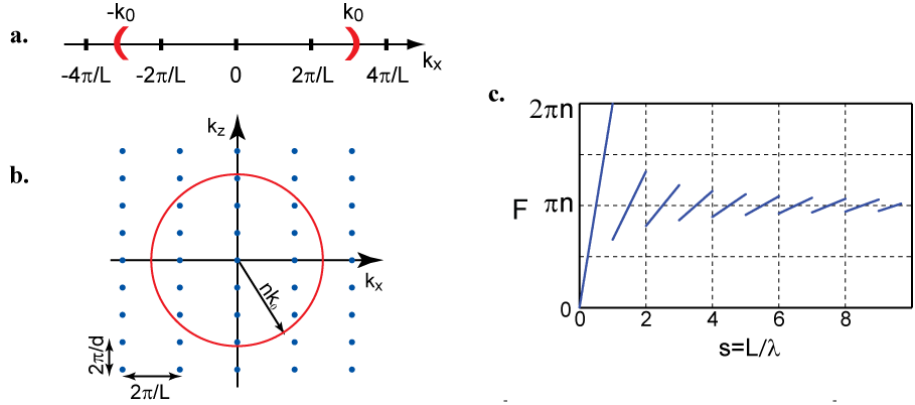

Fig. 2. a) Channels in 1D *k*-space. b) Resonances in a film with 1D grating. Dots represent resonances. c) Upper limit of absorption enhancement in 1D grating films without mirror symmetry.

We first consider the case $L \gg \lambda$. The spacing of the channel is $\frac{2\pi}{L} \ll \frac{2\pi}{\lambda} = k_0$. The discreteness of the channels is therefore not important. The total number of channels (Fig. 2a) at wavelength $\lambda$ thus becomes

$$N = \frac{2k_0}{2\pi/L} = \frac{2L}{\lambda} \tag{8}$$

Notice that we consider only a single polarization. In the frequency range $[\omega, \omega+\Delta\omega]$, the total number of guided resonances supported by the film is (Fig. 2b):

$$M = \frac{2n^2 \pi \omega}{c^2} \left(\frac{L}{2\pi}\right)\left(\frac{d}{2\pi}\right)\Delta\omega \tag{9}$$

Combining Eqs. (7-9), we obtain the upper limit for absorption enhancement

$$F = \frac{A}{d\alpha} = \pi n \tag{10}$$

In contrast to the bulk limit of $4n^2$, the enhancement factor is greatly reduced in structures that are uniform in one of the dimensions. Since the structure considered here is essentially a two dimensional structure, we refer to this limit as the *2D bulk limit*.

In the case when the periodicity is close to the wavelength, the discreteness of the channels becomes important and Eq. (8) is no longer valid. Instead, as shown in Fig. 2a, the number of channels is

$$N = 2\left\lfloor \frac{k_0}{2\pi/L} \right\rfloor + 1 = 2\left\lfloor \frac{L}{\lambda} \right\rfloor + 1 \tag{11}$$

where $\lfloor x \rfloor$ represents the largest integer that is smaller than *x*. We further assume the medium has a high refractive index such that the following conditions are satisfied:

$$L \gg \lambda / n$$
$$d \gg \lambda / n \tag{12}$$

Under these conditions, the resonance in the film can still be approximated as forming a continuum of states, and Eq. (9) is still applicable. The upper limit for the enhancement factor is thus calculated from Eqs. (7), (9) and (11). In Fig. 2c, we plot such an upper limit as a function of normalized frequency $s \equiv L/\lambda$. At low frequency when $L/\lambda < 1$, there is only one channel, i.e. $N = 1$, while the number of resonances increases linearly with frequency. Hence the enhancement factor increases linearly with frequency, reaching its maximum value of $2\pi n$ at $L = \lambda$. At the frequency immediately above $s = 1$, the number of channels increases to $N = 3$, leading to a step-function drop of the enhancement factor. In general, such a sharp drop in enhancement factor occurs whenever new channels appear, i.e. whenever $L = m\lambda$, where $m$ is an integer. Also, in between such sharp drops, the enhancement factor always increases as a function of frequency. In the limit of $L \gg \lambda$, the enhancement factor converges to the 2D bulk limit of $\pi n$.

To test this analytical result, we numerically simulate the 1D grating structure shown in Fig. 1a. We calculate the average enhancement factor $\overline{f} = \int_{\omega_1}^{\omega_2} f(\omega)d\omega/(\omega_2 - \omega_1)$ over the wavelength range from 600nm to 1200nm, which covers the frequency range of $0.5 < s < 1$ in Fig. 2c. $\overline{f}$ is approximately 15 for normally incident light (Fig. 1b,c), which is significantly higher than the 2D bulk limit of $\pi n = 11.1$. As a comparison, using Fig. 2c, the upper limit of the enhancement factor, when averaged over the same frequency range is $\overline{F} = 16.7$. The simulation result therefore compares quite well with the theoretical prediction.

*4.2 Light-trapping limit in symmetrical grating structures*

We now analyze the effects of the symmetry of the grating profile on the light-trapping limit. In contrast to the asymmetrical grating profile shown in Fig. 1a, a symmetrical grating has mirror symmetry in the *x*-direction (Fig. 3a). Ref. [25] proposes to use asymmetrical gratings to reduce the reflection in the normal direction, and thus increase the absorption. On the other hand, the semi-analytical method in Ref. [8] finds no difference for gratings with different symmetries. Here, we analyze the effect of structure symmetry on the fundamental limit of light absorption enhancement with our rigorous electromagnetic approach.

We first discuss the case of normally incident light. Due to mirror symmetry of the film, resonant modes either have an odd or even modal amplitude profile. The normally incident plane wave, which has even modal amplitude profile, cannot couple to modes with odd profiles. Therefore, for the symmetric case, the number of resonances that can contribute to the absorption is reduced by half when compared to the asymmetric case, i.e.:

$$M_{sym} = M/2 \tag{13}$$

where *M* is given by Eq. (9).

In the case where the period is smaller than the wavelength, there is only one channel $N_{sym} = N = 1$. Thus, when the period of the grating is less than the free space wavelength, the symmetric case has a lower enhancement limit $F_{sym} = F/2$ (Fig. 3b).

When the period of the grating is larger than the wavelength, there are more channels, and we also need to consider the effect of symmetry on channels. Due to mirror symmetry, channels can be arranged as even and odd according to

$$S_{even} = \frac{1}{\sqrt{2}}(S_{k_{//}} + S_{-k_{//}})$$
$$S_{odd} = \frac{1}{\sqrt{2}}(S_{k_{//}} - S_{-k_{//}}) \tag{14}$$

Since the incident plane wave from the normal incidence has an even modal amplitude profile, only the even resonances can be excited, which only leak into even channels. Therefore, the number of channels available is also reduced by half $N_{sym} = N/2$ when $L \gg \lambda$. In this case, as seen in Eq. (7), we have the same enhancement limit as that of asymmetrical gratings $F_{sym} = F$ (Fig. 3b). Thus, the symmetry of the grating is important only when the periodicity of the grating is smaller or comparable to the wavelength of incident light. In structures with period much larger compared with the wavelength, the symmetry of the grating does not play a role in determining the upper limit of absorption enhancement.

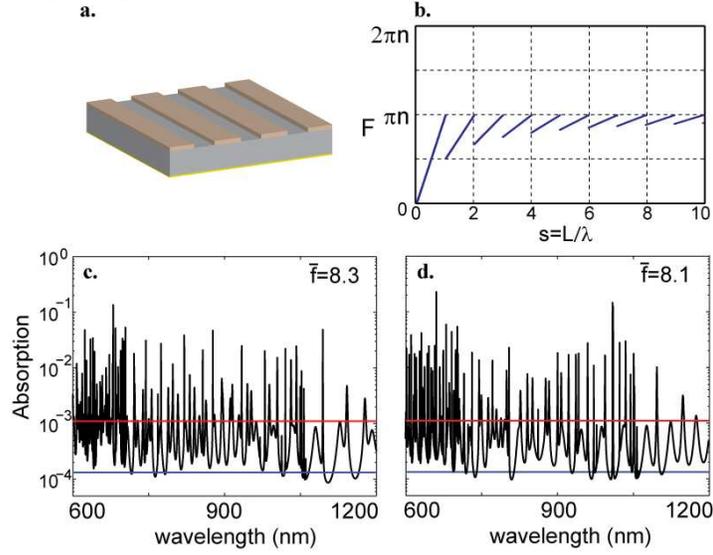

Fig. 3. a) 1D grating structure with mirror symmetry. Air slot width is 40% of the period. The film is the same as that in Fig. 1a. b) Limit of absorption enhancement in grating structures with mirror symmetry. c) and d) Absorption spectra obtained by numerical simulations for s and p polarization lights, respectively. Lights are incident from the normal direction.

For off-normal incident light, the limit of absorption enhancement is the same for symmetrical and asymmetrical gratings, since the incident light no longer has a symmetric modal profile.

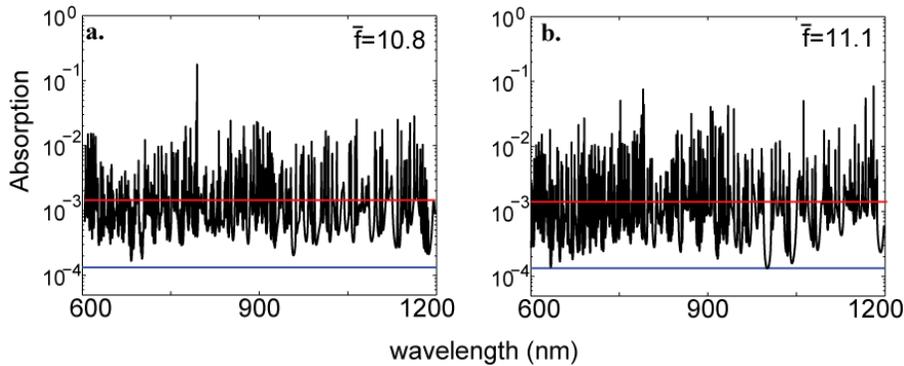

Fig. 4. Absorption spectra for off-normally incident light (averaged among s and p lights). Light incidence angle is 30° and incident plane is oriented at azimuthal angle 30°. a) for the asymmetrical structure in Fig. 1a. b) for the symmetrical structure in Fig. 3a.

To test the analysis above, we simulate the symmetric 1D grating shown in Fig. 3a, where the grating period of 600 nm is chosen such that it is smaller compared to the wavelength range of interest of 600-1200nm. We consider the case of normally incident light first. Theoretically, such a symmetric grating structure should have a spectrally-averaged upper limit of the enhancement factor of $\overline{F} = 8.4$, which is half that of the spectrally-averaged upper limit for the asymmetric case. Numerically, the average enhancement factor $\overline{f}$ is about 8-fold for normally incident light (Fig. 3c and Fig. 3d), which compares well with the theoretical prediction, and is indeed far lower compared with the asymmetric case shown in Fig. 1. Finally, while such a structure behaves differently for the two different polarizations, as seen in the spectra plotted in Figs. 3c and 3d, the theoretical upper limit for absorption enhancement is in fact the same for both polarizations. Numerically, the spectrally averaged enhanced factors for the two polarizations are also quite similar (Fig. 3c and d.)

We have also simulated the response of the symmetric grating structure of Fig. 3a for off-normal incident light at an angle of incidence of 30-degrees. (Fig. 4a). The spectrally-averaged enhancement factor of 11-fold is significantly higher compared with normal incident light for such a symmetric grating structure, consistent with the theoretical discussion above. Moreover, at the same angle of incidence of 30-degrees, the asymmetric grating structure of Fig. 1a also shows a spectrally averaged enhancement factor of 11-fold (Fig. 4b). Thus, the symmetry of the grating indeed does not play a significant role for off-normal incident light as far as the spectrally averaged enhancement factor is concerned.

## 5. 2D grating

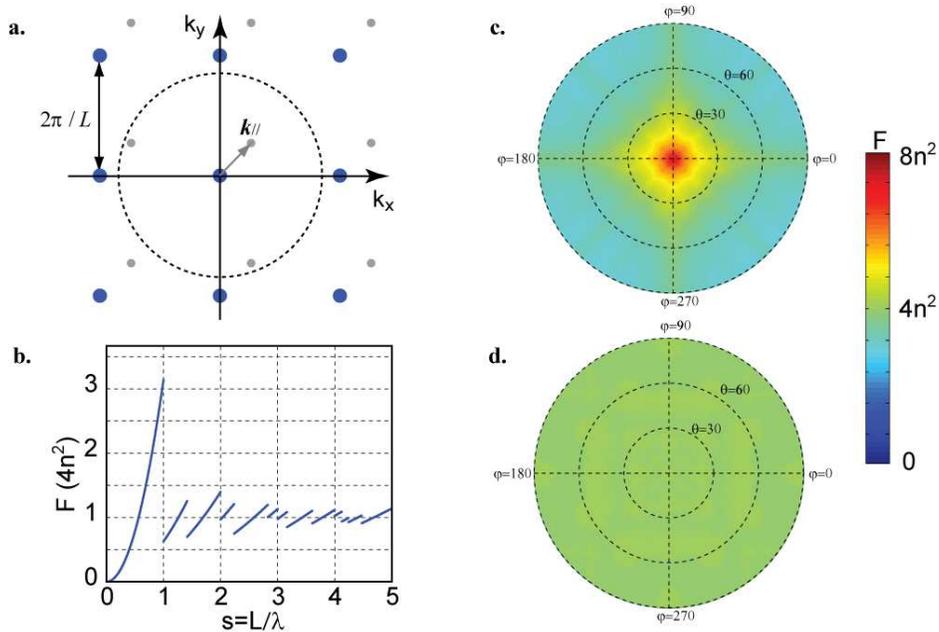

Fig. 5. a) Channels in 2D $k$-space. Blue and grey dots represent channels when the incident light comes in from normal and off-normal directions, respectively. The radius of the circle is $k_0$. b) Limit of absorption enhancement in 2D grating structures. c) Angular response of the average enhancement factor integrated over wavelength from 600nm to 1200nm. The periodicity of the grating is 600nm. d) the same as (c) except the periodicity is 3000nm.

The upper limit in a 1D grating structure falls far below the conventional bulk limit of $4n^2$ due to a reduced number of resonances. To achieve a higher enhancement factor, it is always better to use 2D gratings that are periodic in both $x$ and $y$ directions, since a 2D grating allows access to all resonances supported by a film. In this section, we present a detailed discussion on the upper limit of light trapping with a 2D grating based on Eq. (7).

*5.1. Theoretical upper-limit of enhancement factor for 2D grating structures*

We first consider the case where the grating has a square lattice, with a periodicity of $L$ in both dimensions. In the frequency range $[\omega, \omega+\Delta\omega]$, assuming Eq. (12) is satisfied, the total number of guided resonances supported by the film is

$$M = \frac{8\pi n^3 \omega^2}{c^3} \left(\frac{L}{2\pi}\right)^2 \left(\frac{d}{2\pi}\right) \Delta\omega. \tag{15}$$

Notice that the number of resonances increases quadratically as a function of frequency.

When a plane wave is normally incident upon the film, the grating can excite plane waves in other propagating directions in free space. The parallel wavevectors $G_{m,n}$ of these excited plane waves in free space are:

$$G_{m,n} = m\frac{2\pi}{L}\hat{x} + n\frac{2\pi}{L}\hat{y}, \tag{16}$$

with $m$ and $n$ being integers. Hence these parallel wavevectors form a square lattice in the wavevector vector space (Fig. 4a, blue dots). Moreover, since these are propagating plane waves, their wavevectors need to lie within a circle as defined by $|G| < k_0 \equiv \frac{\omega}{c}$ (Fig. 5a, dashed line). The total number of different wavevector points, multiplied by two in order to take into account both polarizations, defines the number of channels $N$ that is required for the calculation using Eq. (7)

At low frequency when $s = L/\lambda < 1$, there are only two channels ($N$=2), accounting for two polarizations, in the normal direction, while the number of resonances increases with frequency quadratically. Hence the enhancement factor increases quadratically with frequency and reaches its maximum value of $4\pi n^2$ at $s$=1 (Fig. 5b). Immediately above $s=1$, the wavelength becomes shorter than the period $L$. The number of channels increases to $N$=10, leading to a step-function drop of the enhancement factor. In the limit of $L \gg \lambda$, the calculated upper limit reproduces the 3D bulk limit of $4n^2$ [9].

Next, we analyze the angular response of the upper-limit of the enhancement factor by considering a plane wave incident from a direction specified by an incidence angle $\theta$ and an azimuthal angle $\varphi$. Such an incident plane wave has a parallel wavevector $k_{//} = k_0 \sin(\theta)\cos(\varphi)\hat{x} + k_0 \sin(\theta)\sin(\varphi)\hat{y}$. In the presence of the grating, such a plane wave can excite other plane waves in free space with parallel wavevectors:

$$k = k_{//} + G_{m,n}, \tag{17}$$

where $G_{m,n}$ is defined in Eq. (16), provided that these parallel wavevectors are located in a circle in k-space $|k| < k_0$, as is required for propagating plane waves.

In comparison to the normal incident case described by Eq. (16), Eq. (17) defines a similar square lattice in the wavevector space, except that the positions of the lattice points in $k$-space are now shifted by the parallel wavevector $k_{//}$ of the incident plane wave (Fig. 5a). As a result of such a shift, one can see that the number of channels $N$, which is proportional to the number of $k$-space lattice points that fall within the circle of $|k| < k_0$, is dependent on the incidence angle. Such angular dependency is particularly strong for the case where there are only small numbers of channels, which occurs when the periodicity is about wavelength scale.

As an example, Fig. 5a illustrates the case where the normalized frequency $s = L/\lambda = 0.87 < 1$. For normal incidence, there is only one $k$-space lattice point (as shown by the blue dot in Fig. 5a) within the circle of $|k| < k_0$ (dashed line in Fig. 5a), corresponding to two channels. At the same frequency, for off-normal incident light with $k_{//} = 0.3k_0 \hat{x} + 0.3k_0 \hat{y}$, corresponding to a plane wave incident from a direction as defined by $\theta = 25°$, $\varphi=45°$, there are 3 $k$-space lattice points within the circle of $|k| < k_0$ (grey dots in Fig. 5a), and hence 6 available channels. Thus, for a plane wave incident from such an off-normal incident direction, the upper limit for absorption enhancement should be only 1/3 of the upper limit for normally incident light. This example illustrates that there can be substantial angular dependency in absorption enhancement when the periodicity is comparable to the wavelength. In contrast, when the periodicity is much larger than the wavelength, the $k$-space lattice points are densely distributed and the number of channels $N \gg 1$. Therefore, the shift of the lattice points in $k$-space due to different incidence angles has negligible effect on the total number of channels. In the case where $L \gg \lambda$, the upper limit of enhancement is not sensitive to the incidence angles.

In Fig. 5b, we provide detailed analytic results regarding the angular dependency of the upper limit of absorption enhancement. We consider a grating with a period $L = 600$ nm. For each direction of incidence as specified by an angle of incidence $\theta$ and an azimuthal angle of $\varphi$, we calculate the upper limit $F$ at each frequency using Eq. (7). We then average the upper limit $F$ calculated over the wavelength range between 600nm and 1200nm, and plot the spectral average $\bar{F}$ as a function of $\theta$ and $\varphi$ in Fig. 5c. In this case, the grating period is smaller than the wavelength of interest. In the vicinity of the normal direction, $\bar{F}$ is approximately $8n^2$. Thus, it is possible to use a grating structure to obtain broad-band enhancement of absorption higher than $4n^2$ (Fig. 5c). This result is consistent with Fig. 5b: the wavelength range between 600 and 1200nm corresponds to the range of $0.5 \le s \le 1$, where the enhancement factor is above $4n^2$. However, when the incident light deviates from normal direction, $\bar{F}$ starts to drops (Fig. 5c). For incidence angles larger than $60°$, $\bar{F}$ drops well below $4n^2$, showing a strong angular dependency.

As a second example, we consider the case where the period $L = 3000$ nm, and the same wavelength range of 600 to 1200nm, which for this grating periodicity corresponds to the normalized frequency range $2.5 \le s \le 5$. In this case, the period is considerably larger than the wavelength, and the number of channels is much larger than 1. The spectrally averaged upper-limit of enhancement has much weaker angular dependency. As shown in Fig. 5d, $\bar{F}$ is around $4n^2$ for all incidence angles, showing a near-isotropic response.

To conclude this section on the theoretical analysis of 2D grating structures, we note that a grating with wavelength scale periodicity can achieve an absorption enhancement factor that is higher than $4n^2$ over a broad-range of wavelengths. Such enhancement, however, comes at the expense of substantial angular dependency. As a result, a grating structure in general, when the film thickness is a few wavelength thick, cannot overcome the conventional bulk limit of $4n^2 / \sin^2 \theta$ [4]. On the other hand, a grating structure with periodicity much larger than the wavelength has enhancement factor approaching $4n^2$ with near-isotropic response.

*5.2 Numerical simulations of 2D grating structures*

In this section, we use numerical simulations to support the analytical predictions made in Section 5.1. In our numerical simulation, we consider the 2D grating structure whose unit cell shown in Fig. 6a. The periods of the grating are 600nm in both $x$ and $y$ directions. The film structure otherwise is the same as in Section 2.

For normally incident light, the absorption spectrum in the wavelength range 600-1200nm is shown in Fig. 6b. The spectrally averaged enhancement factor in this wavelength range is $\bar{f} = 74.2$, which significantly exceeds $4n^2 = 50$. This is also much larger than that of 1D grating with the same period (Fig. 1) for the same wavelength range. As a comparison, the spectrally averaged theoretical upper limit is $\bar{F} = 91$ as calculated from Fig. 5b. The numerical result is in good agreement with our theoretical prediction. The simulated enhancement factor is lower than the theoretical upper limit, due to the fact that some resonances are not in the over-coupling regime.

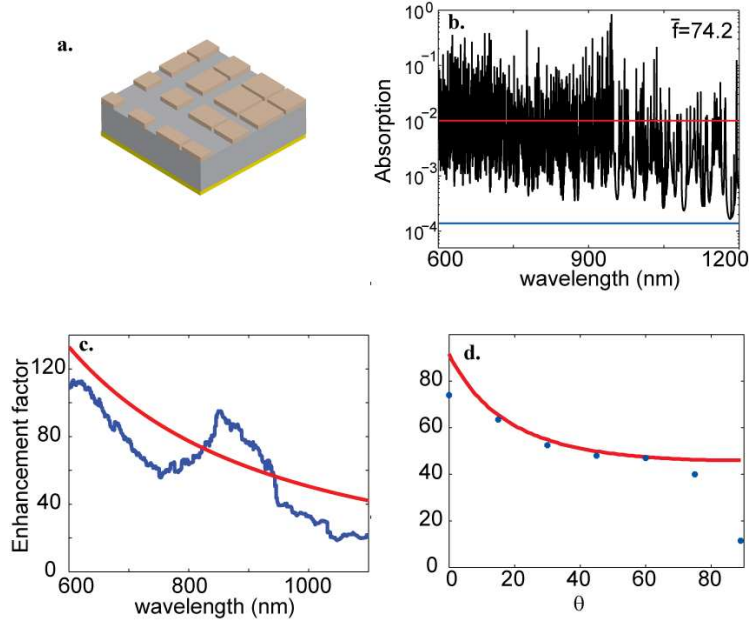

Fig. 6. a) The unit of cell of a 2D grating structure. Film thickness is the same as the structure in Fig. 1a. b) Absorption spectrum for normally incident light (averaged among two polarizations). c) Spectra of the running average enhancement factors. Blue: simulation result. Red: analytical theory. d. average enhancement factor for different incidence angles $\theta$. The azimuthal angle is fixed at $\varphi = 0$. Blue dots are simulation results and red is analytical theory.

In the absorption spectrum of Fig. 6b, the higher frequency region has denser peaks, since the density of resonances in the film in general increases with frequency. In our theoretical analysis, the enhancement factor is directly proportional to the number of resonances. Thus, one should expect that in the higher frequency region the enhancement factor is larger. To study this effect, we calculate the running average of the enhancement factor over a bandwidth of 100nm. The enhancement factor indeed decreases with increasing wavelength (Fig. 6c). As a comparison, we also calculate the running average for the theoretical upper limit as shown by red line in Fig. 6c, which shows good agreement with numerical results.

The theoretical curve in Fig. 6c is generated using Eq. (15), where the spectral distribution of the resonances is calculated from the three-dimensional density of states in a bulk medium. Such a calculation is strictly speaking only valid for films that are sufficiently thick. In our simulated structure, at a film thickness of 3 microns, there is still a finite-size effect. As a result one should expect some deviations between the simulation and the theory. The agreement between the simulation and theoretical curves is expected to be better as the thickness of the film increases.

To investigate the angular response of the grating, we numerically calculate the spectrally-averaged enhancement factor $\bar{f}$ as a function of the incidence angle $\theta$. To simplify the calculation, we fix the azimuthal angle at $\varphi = 0$. The numerically obtained enhancement factor exceeds $4n^2$ at normal incidence, (Fig. 6d), and decreases as the angle of incidence increases, consistent with the theoretical prediction (red line Fig. 6d). The numerically obtained enhancement factor is below, but fairly close to the theoretical maximum, indicating that the structure shown in Fig. 6 is in fact already quite optimal over a large range of incidence angles. At very large incidence angles ($\theta > 80°$) the numerical simulation shows much lower enhancement compared to theory. At these angles, the coupling between incident light and resonances become much weaker, and resonances are no longer in the over-coupling limit.

*5.3 Influence of periodic lattice structure*

Finally, we briefly comment on the influence of the lattice periodicity of the grating structure. The analysis above has focused on square lattices. In practice, a triangular lattice is often found in closely packed nano-particles and nano-wires [37]. For a grating with a triangular lattice with period $L$, the channels form a triangular lattice in $k$-space (Fig. 7a), The distance between the channel at the origin of k-space and its nearest neighbors is $k_T = (2/\sqrt{3})2\pi/L$, which is larger than that of the square lattice $k_S = 2\pi/L$. As a result, for normally incident light, the frequency range where the grating operates with only 2 channels is larger ($0 < s < 2/\sqrt{3}$) compared to the case of the square lattice ($0 < s < 1$). This leads to a higher maximum enhancement factor $8n^2\pi/\sqrt{3}$ (Fig. 7b). On the other hand, as the period becomes larger, the maximum enhancement ratios in gratings with different lattices all converge to the same bulk limit of $4n^2$.

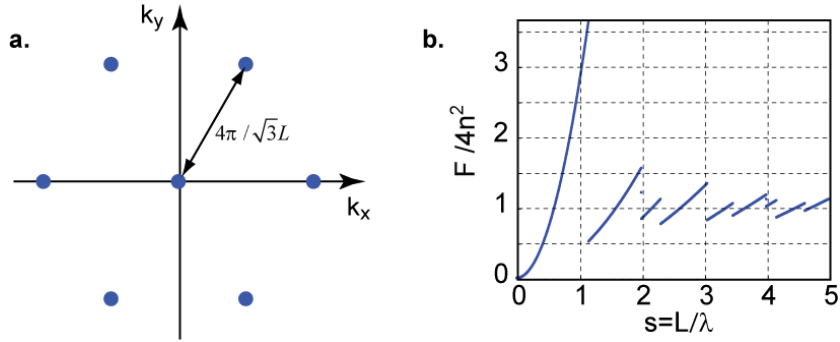

Fig. 7. a) Channels in 2D *k*-space for a grating with triangular lattice periodicity. The lattice constant is *L*. b) Upper limit of absorption enhancement for gratings with triangular lattice periodicity.

## 6. Conclusion and Outlook

In conclusion, we analyzed the limit of light-trapping enhancement in grating structures using statistical coupled-mode theory. The presence of gratings modifies the channels available in free-space, particularly when the periodicity is about wavelength-scale. An enhancement factor beyond $4n^2$ can be achieved, but with significant angular response. We also show that 2D gratings have a higher enhancement factor compared to 1D gratings. In addition, we show the effects on absorption enhancement of the symmetry of a grating profile.

The conclusions of our paper are applicable for films that are at least a few wavelengths thick, and have weak absorption. The theoretical formalism of Section 3, however, has more general applicability. In particular, such a theory can be used to study films having thicknesses comparable to, or even smaller than, the wavelength of light [Ref. 9].

The theory developed here is for an idealized situation where the absorption of the film is assumed to be weak over a very wide range of frequencies. Such an idealized situation is important for understanding the ultimate limit of light absorption enhancement. One should note, however, that absorption enhancement in realistic material systems can be considerably more complex. The dashed line in Fig. 8, for example, shows the absorption spectrum of silicon, for a film thickness of 3 micron. Notice that the absorption varies very significantly over the wavelength range of 600-1200nm. The solid line in Fig. 8 shows the absorption spectrum, using the same grating structure shown in Fig. 6a placed on top of such a silicon film. Since the film is no longer in the weak-absorbing limit, especially in the shorter wavelength range, the absorption enhancement ratio is far below the theoretical limit derived for the weak absorber. However, we notice that the grating structure as shown in Fig. 6a in fact provides very significant absorption enhancement over the entire wavelength range considered here. Thus the grating structure considered is practically important. Understanding the upper limit of absorption enhancement in nanophotonic structures, over the entire absorption bandwidth of a given material where the material absorption can vary by many orders of magnitude, is certainly a very important future direction.

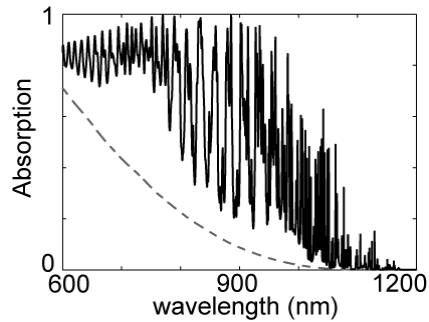

Fig. 8. Absorption spectrum for structure shown in Fig. 6a. except that the film is replaced with crystalline silicon. Dashed line is single-pass absorption spectrum.

## Acknowledgements

The work was supported by DOE grant No. DE-FG02-07ER46426.